\documentclass[a4paper]{llncs}

\usepackage{amssymb}
\usepackage{graphicx}
\usepackage{tabularx}
\usepackage{tabulary}
\usepackage{booktabs}
\usepackage{url}
\usepackage{hyperref}
\usepackage[all]{hypcap}
\usepackage{enumitem}
\usepackage{todonotes}
\usepackage{xspace}
\usepackage{listliketab}
\usepackage{lipsum} 
\usepackage{url}
\usepackage[numbers]{natbib}

\newcommand{\GermanWeb}{Popular German Web\xspace}

\begin{document}

\mainmatter  

\title{Semantic URL Analytics to Support Efficient Annotation of Large
Scale Web Archives}


%
%
\author{Tarcisio Souza$^1$ \and Elena Demidova$^1$ \and Thomas Risse$^1$ \and Helge Holzmann$^1$ 
\and \\Gerhard Gossen$^1$ \and Julian Szymanski$^2$}

%

\institute{$^1$L3S Research Center and Leibniz Universit\"at Hannover\\
Hannover, Germany\\
name@L3S.de\\
$^2$Gdansk University of Technology\\
Poland\\
julian.szymanski@eti.pg.gda.pl\\
}


%
%

\maketitle

\begin{abstract}

Long-term Web archives comprise Web documents gathered over longer time periods and can easily reach hundreds of terabytes in size.
Semantic annotations such as named entities 
can facilitate intelligent access to the Web archive data. 
However, the annotation of the entire archive content on this scale is often infeasible. 
The most efficient way to access the documents within Web archives is provided through their URLs, which are typically stored in 
dedicated index files.
The URLs of the archived Web documents can contain semantic information and can offer an 
efficient way to obtain initial semantic annotations for the archived documents. 
In this paper, we analyse the applicability of semantic analysis techniques such as named entity extraction
to the URLs in a Web archive. We evaluate
the precision of the named entity extraction from the URLs in the \GermanWeb dataset
and analyse the proportion of the archived URLs 
from 1,444 popular domains in the time interval from 2000 to 2012 to which these techniques are applicable.
Our results demonstrate that named entity recognition can be successfully applied to a large number of URLs in 
our Web archive and provide a good starting point to efficiently annotate large scale collections of Web documents.
\end{abstract}

\section{Introduction and Motivation}
\label{sec:intro}

Web archives are a unique source of data reflecting rapid evolution of the digital world.
Recently, an increasing interest in using the data stored within the Web archives for research purposes has been observed in several disciplines such as history or digital sociology \cite{icrawlRequirements}. For example, as discussed in \cite{Brugger14}, Web archives can be an important source for communication and media history and within historiography in general.

Unfortunately, existing Web archives are very difficult to use since most often only a URL based access is provided. Researchers are typically interested in a few relevant Web sites regarding a given topic, domain or time frame to be selected for manual analysis. Finding such relevant data for a specific purpose within the Web archive is still very challenging. This is mainly attributed to the very large size of the data coupled with the lack of efficient tools for annotation, search and exploration. Indexing existing Web archives, which often contain hundreds of terabytes of data, is difficult and hence, full-text search capabilities are rarely available, not to mention more sophisticated semantic content analytics.

In this context, we look at different ways to obtain relevant documents from a Web archive efficiently and analyse the role of URLs of the archived documents towards this goal. The advantage of using URLs is twofold: First, the URLs of the archived documents can be retrieved from the archive efficiently, using CDX files (i.e. standard index files that contain URLs and additional metadata, such as mime type and capture dates). Second, URLs can contain important hints about the document content. In this context, related work on URL analytics on the Web shows that URLs can provide accurate estimates of the document language \cite{Baykan:2008}, location relevance \cite{Anastacio:2009} and topic classification \cite{URLFeatures12}.

In this paper we analyse the applicability and precision of Named Entity Extraction (NER) in the context of URLs. Furthermore, we analyse the distribution of the extracted named entities within the URLs in the \GermanWeb dataset - a subset of the Internet Archive data covering popular domains in the ``.de'' top level domain over a period of 12 years. This analysis confirms the precision of NER in the context of URL analytics and helps to better understand which domains can be efficiently accessed using such light-weight annotation methods for a Web archive.
Our results demonstrate that state-of-the-art NER tools, such as 
Stanford NER\footnote{http://nlp.stanford.edu/software/CRF-NER.shtml}, can achieve high precision 
(up to 85\% in our dataset) if applied to the URLs after performing sufficient
pre-processing and post-filtering described in this paper.
We also observed that the number of extracted entities differs significantly across the domain 
categories and along the temporal dimension. In some years, the dataset contains dominant 
domains - i.e. the domains within a domain category that contribute the majority of captures in 
the specific year. In most cases, the variations in the extraction results can be explained by 
the varying number of captures from such dominant domains as well as by the entity-rich URLs 
in such domains (i.e. the domains like \textit{dblp.uni-trier.de} - an open
computer science bibliography, where the URLs typically contain named entities
of the type person, \textit{dict.tu-chemnitz.de} - a dictionary domain, that
frequently presents entities of miscellaneous type).

Overall, our results confirm that NER is a useful method of semantic URL analytics and can provide precise results and high coverage in several domain categories.


\section{The \GermanWeb: a Dataset Description}
\label{sec:dataset}

The dataset used in this study is referred to as ``\GermanWeb''. This dataset is a subset of the '.de' top-level domain (tld), 
as it has been archived by the Internet
Archive\footnote{\url{http://archive.org}} and provided to us in the context of
the ALEXANDRIA\footnote{\url{http://alexandria-project.eu/}} project.
This dataset comprises the most prominent domains in 17 categories from 2000 to 2012 selected according to the 
Alexa ranking\footnote{\url{http://www.alexa.com}}.

\textbf{Terminology:} A \textit{URL (uniform resource locator)} identifies a Web resource (for example a Web page) 
and specifies its location on the Web. Over time, the content of the Web page under any given URL may change.
Therefore, Web archives often re-assessed the URLs after a period of time to
collect new content.
In the following we refer to a particular copy of the URL assessed at a certain time and stored in the archive as
a \textit{capture}. This way, a Web archive can possess several captures of the URL, whereas each capture 
can be uniquely identified through the URL and the date.

\begin{figure}[h]
\vspace*{-5mm}

  \caption{CDX structure}
  \centering
    \includegraphics[width=0.8\textwidth]{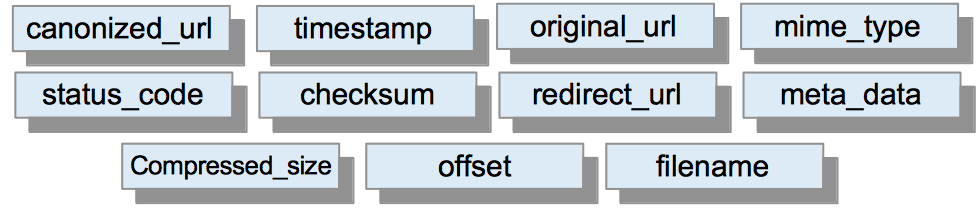}
    
    \label{fig:cdx}
    \vspace*{-12mm}
    
\end{figure}

These captures are stored as CDX files that contain meta information about the
crawls in a space-separated format with one line per capture, i.e.
one snapshot of one URL at a given time. The corresponding line in the CDX file
presents the structure illustrated by Fig.~\ref{fig:cdx}. We consider
``original url'' as the input for NER, ``timestamp'' provides the exact
time when the URL was crawled and ``status code'' tells us whether or not a
successful response was returned. The other fields are ignored since they do not represent useful
information for our analysis.	

\subsection{Domain Extraction}
There is an interest in these domains as they are impacting most users of
the Web today and at the same time they have the biggest impact on upcoming research. 
The selection of domains was taken from different categories on Amazon's Alexa. 
In order to match our dataset we fetched only websites from those categories on that comprise 
German websites\footnote{\url{http://www.alexa.com/topsites/category/Top/World/Deutsch}}. In addition to the 15 top categories, 
we picked \textit{news} and \textit{universities} as two sub-categories that
seemed particularly relevant to analyze separately in our study.

As the available dataset only included domains under the German tld '.de', we filtered out all websites on Alexa's list that are German 
but use a different tld. 
Out of the remaining, we extracted up to 100 from the top of every category. 
The final state of this ranking, which is analyzed here, was retrieved on July 10th, 2014
at 09:26 CET.


\subsection{Dataset Cleaning and Pre-Processing}
\label{sec:pipeline}



Following the domain extraction we performed a few cleaning steps at the URL level to filter out malformed URLs 
as well as those that are inappropriate for our analysis:
\begin{itemize}[noitemsep,nolistsep]
\item In this analysis we focus on the html content of the Web archive. 
Therefore, we discarded all captures that do not represent the html content (identified by .html or .htm extensions in the URLs).
\item We discarded all captures of the URLs that never returned a
successful status code (i.e. a status code starting with ``2'', according to the official HTTP status codes).


\end{itemize}

\begin{figure}[htb!]
\vspace*{-8mm}

  \caption{URL processing pipeline}
  \centering
    \includegraphics[width=1\textwidth]{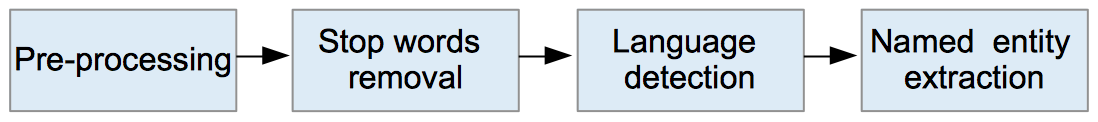}
    \label{fig:pipeline}
    \vspace*{-12mm}
    
\end{figure}

Fig.~\ref{fig:pipeline} illustrates the sequence of steps applied to pre-process the URLs,
determine their language and, finally, to extract semantic information such as
named entities. Regarding the URL tokenization, we considered only words
extracted from the URL string also excluding special characters and numbers, based on 
a simple regular expression.

\begin{itemize}
  \item \textbf{Pre-processing:} From the pre-processing perspective, we
  consider only the URL path, discarding the extension. We exclude other parts such as domain name,
parameters and numbers (e.g. port number, session id, etc.) from further consideration.
These fields are not expected to contain semantic information specific to the
particular Web document. 
  \item \textbf{Stop words removal:} In next step we detect stop words within the
parsed URLs and eliminate them. In order to determine stop words, we randomly
sample URLs and manually select the most frequent terms extracted from that
sample. 
This is required, as no pre-defined stop word lists for URLs exists. Short terms (i.e. the terms with the length less than
three characters) are also discarded in this step. Stop words filtering is particularly important to 
increase precision of the language detection performed in the next step.
  \item \textbf{Language detection:} We apply an n-gram language detection model to 
the remaining tokens in the URLs to detect the language of the URL and to select the 
correct language configuration for the extractors applied in the next step. 
  \item \textbf{Named entity extraction:} We apply Stanford NER to extract named entities mentioned in the preprocessed URLs.
  The language-specific named entity extractor (German or English) is selected based on the determined URL language. 
\end{itemize}
%




\subsection{Dataset Statistics}
\label{sec:statistics}

Ultimately, we obtained a dataset consisting of 17 categories with today's popular domains 
from the '.de' top level domain, 
as presented in Table~\ref{tab:dataset_statistics}. The resulting \GermanWeb dataset covers 
1,444 domains with more than 320 million captures in total. Table~\ref{tab:dataset_statistics} presents the number of 
domains and captures in each domain category as well as the percentage of captures from which we could extract named entities using 
our method. As we can observe, the highest coverage of captures containing
entities is attributed to the \textit{education} category with 73\%, followed by
the \textit{regional} and \textit{sports} categories with around 40\%.
The dominating domains for the \textit{education} category 
were \textit{stayfriends.de} and \textit{wer-weiss-was.de}, with 20-50\% of
captures dependent on the year.
These domains typically contain entity-rich URLs that explains the high percentage of entities extracted from this category.
For many other domain categories the coverage of captures containing named entities exceeds 20\%.
Overall, we can say that our method can efficiently produce annotations for a significant number of captures in many domain categories. 



\begin{table}[!t]
\vspace*{-5mm}
  \small
  \centering  
  \caption{The \lowercase{\GermanWeb} dataset details: the number of domains and
  captures per category.}
\begin{tabular}{rrrr}
    \toprule
    \textbf{Category}&\textbf{\#~Domains}&\textbf{\#~Captures}&\textbf{Entities(\%)}\\
    \midrule
    Education & 100 & 12,406,130 & 73.36\\
    Regional & 100 & 34,204,862 & 44.79\\
    Sports & 100 & 17,358,130 & 39.33\\
    Business & 100 & 25,457,639 & 36.39\\
    Recreation & 100 & 8,260,029 & 30.95\\
    Media & 100 & 11,277,003 & 28.20\\
    Universities & 100 & 14,299,856 & 25.09\\
    News &  40 & 41,710,500 & 23.13\\
    Shopping & 100 & 33,045,310 & 20.14 \\
    Culture & 100 & 6,822,986 & 19.69\\
    Society & 100 & 9,968,534& 18.37\\
    Games & 99 & 13,518,500 & 16.40\\
    Computer & 100 & 26,298,534 & 15.90\\
    Home & 100 & 45,488,255 & 14.07\\
    Kids~\&~Teens & 10 & 1,682,848 & 10.45\\
    Health & 100 & 6,260,340 & 9.31\\
    Science & 100 & 13,651,913 & 7.86\\
    \hline
    \textbf{\textit{TOTAL}} & \textbf{\textit{1444}} &
    \textbf{\textit{321,711,369}} \\
    \bottomrule
  \end{tabular}
  
  \vspace*{-5mm}
  \label{tab:dataset_statistics}
\end{table}

\begin{figure}[htb!]
    \vspace*{-8mm}

  \caption{Overall distribution of the number of captures per year.}
  \centering
    \includegraphics[width=0.8\textwidth]{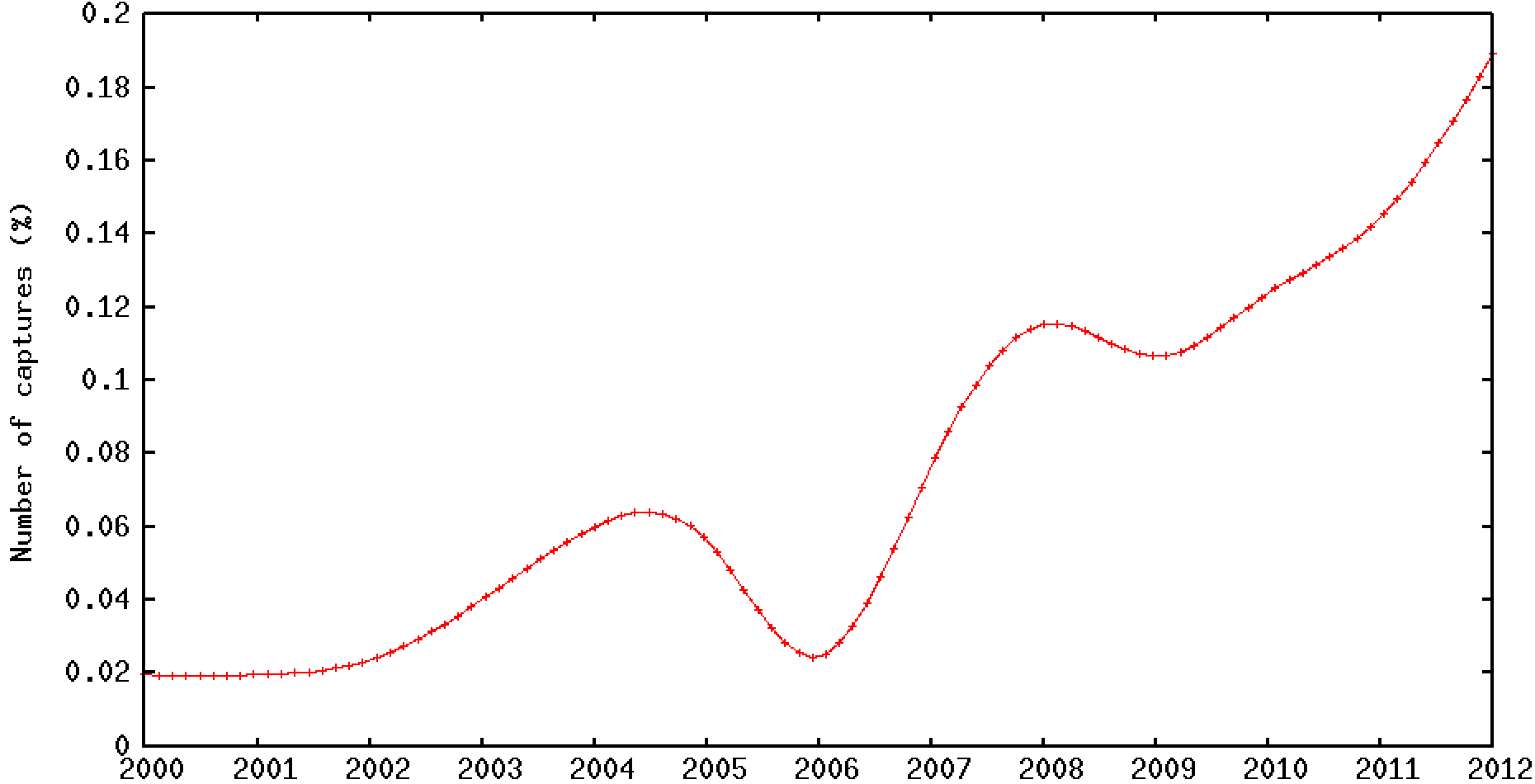}
    \label{fig:captures}
        \vspace*{-5mm}
    
\end{figure}

In order to obtain a better understanding of the temporal dimension of the dataset, we analyse the distribution 
of captures over time. Fig.~\ref{fig:captures} illustrates the overall distribution of the number of captures per year
in the time period from 2000 to 2012 normalized by the total number of captures
in the \GermanWeb dataset, where X axis represent the years and Y axis the
percentage of captures. 

This graph illustrates that the number of captures in
the dataset is rapidly increasing over time, in particular starting from 2007, although at some points (2005 - 2006 and 2009) temporary decreases in the data collection rate can be observed.
The total number of captures are more equally distributed for the majority of
years per domains, where \textit{spiegel.de} appears at
the first position from 2001 to 2012, representing 7.72\% of all captures per year, on
average. The university domain \textit{tu-berlin.de} dominates the crawl by
representing 47.30\% of all captures in 2000.


\begin{figure}[htb!]
\vspace*{-8mm}

  \caption{Number of captures within selected domain categories.}
  \centering
    \includegraphics[width=0.9\textwidth]{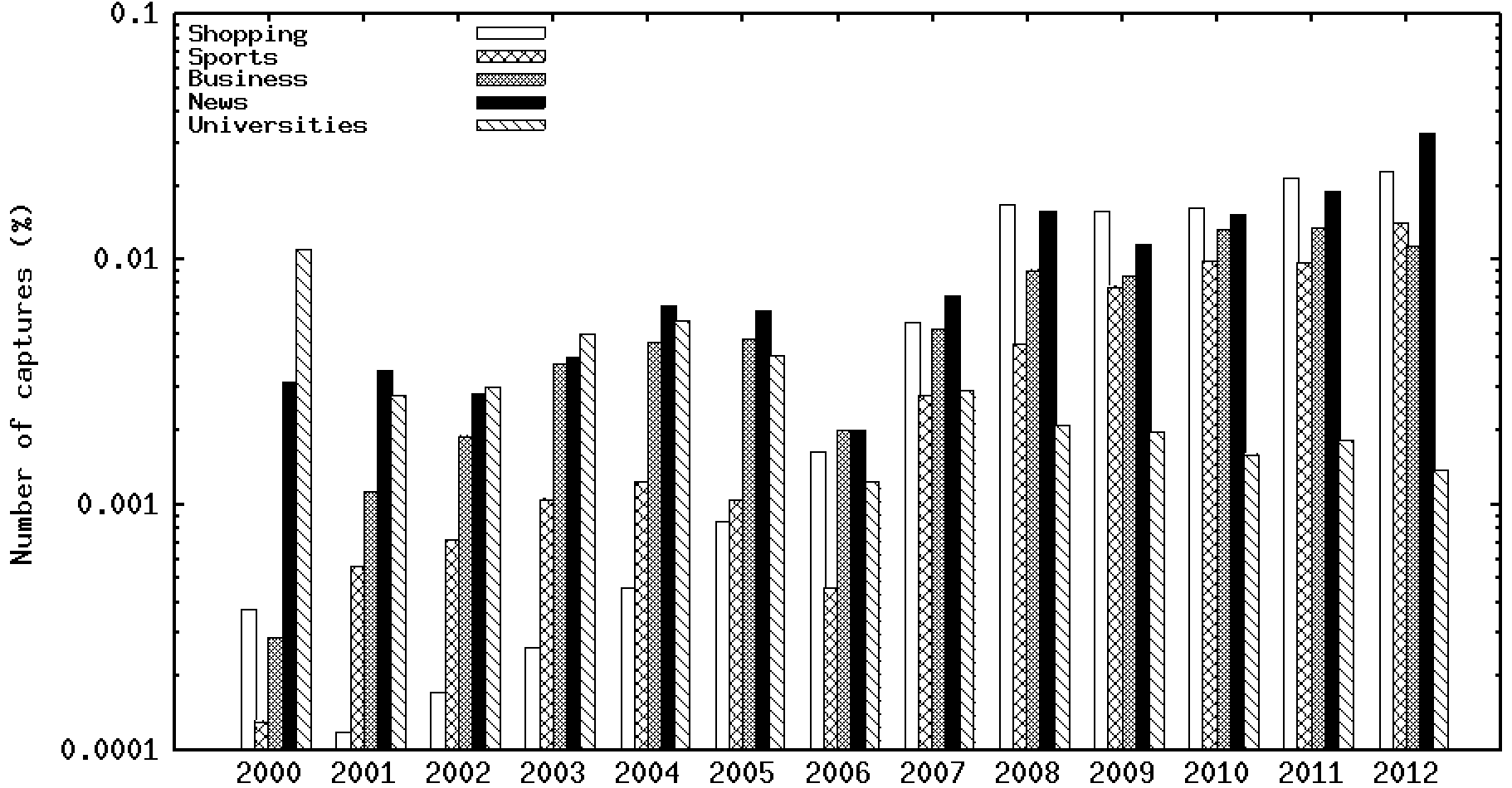}
    \label{fig:captures_domains}
    \vspace*{-6mm}
    
\end{figure}

Fig.~\ref{fig:captures_domains} shows the distribution of all captures through
five selected domain categories including \textit{shopping},  
\textit{sports}, \textit{business}, \textit{news} and  
\textit{universities} normalized by the total number of captures.
As we can observe, various domain categories have 
different dynamics within the archive: 
Whereas the proportion of the \textit{news} cites is relatively high and 
is only slowly increasing over time, the proportion of 
\textit{shopping} cites increased rapidly in recent years.
In contrast, the proportion of the \textit{universities} cites 
is slowly decreasing, in particular starting from 2007.

Even the total number of captures of \textit{spiegel.de} (a german
\textit{news} domain) exceeds total captures from top \textit{universities} domains
(e.g.
\textit{uni-leipzig.de} and \textit{cert.uni-stuttgart.de}) from 2002 to 2003,
the overall number of captures from \textit{universities} is greater than
\textit{news} for those years. In this period the majority of domains belongs to
\textit{universities} (140) and only 40 belongs to \textit{news}.
From 2008 to 2011, captures from \textit{shopping} (532 domains) are more
frequent followed by \textit{news} (136 domains) and \textit{business} (588
domains).

Overall Fig.~\ref{fig:captures_domains} shows the behavior of captures within
five selected categories, where the quantity of captures in general is growing,
except for \textit{universities} starting in 2007. We also observed several
\textit{business} domains from 2008 to 2011, but the majority of captures
belongs to \textit{shopping} in this period.


\section{URL Analytics}
\label{sec:analysis}

In this section we describe the results of 
the URL analysis we obtained by applying the pipeline described in
Section \ref{sec:pipeline} to the \GermanWeb dataset described in Section \ref{sec:statistics}.
The goal of the analysis is two-fold: 
First, we evaluate the precision of the
named entity extraction method for URLs proposed in this paper to confirm its effectiveness; 
Second, we would like to better understand 
the domain coverage and the temporal coverage of the proposed method while applied to our dataset.
In this section we present the evaluation results of the method and statistics we collected while
applying the method to the \GermanWeb dataset.

\subsection{Language Detection Statistics}
\label{sec:analytics-lang}

In order to detect the language of a URL, we applied state-of-the-art techniques to language detection using 
n-grams \cite{Baykan:2008}.
The URL pre-processing described in Section \ref{sec:pipeline}, such as URL splitting and removal of URL-specific stop words 
makes it possible to apply the n-gram analysis on the relevant part of the URL only
and to increase precision of the language detection.
The stop words typical for the URLs are identified using a random sample from the whole URL collection and a manually 
identified frequency threshold.

According to the results of the language detection analysis, 52.89\% of the URLs in our \GermanWeb dataset
are in German, 27.96\% in English and 19.14\% in other languages. After
applying the URL pre-processing, we obtained 89\% of precision for language
detection. We measure this precision choosing a random sample of 100 URLs and
manually checking the returning language.
%
%
%
 
\subsection{Precision of the Named Entity Recognition for URLs}
\label{sec:analytics-ner}

In this section we describe our evaluation results regarding the NER precision applied to the German and the English URLs 
in the \GermanWeb dataset.


State-of-the-art Named Entity Recognition (NER) techniques are language dependent. Therefore, we restrict the NER processing 
to the URLs detected as German and English and apply language-specific configurations of the Stanford NER to these URLs.
With this restriction, we cover more than 80\% of the URLs in the \GermanWeb dataset (s. above). 

In order to evaluate precision of Stanford NER on this dataset, we manually
chose a random sample of 100 URLs out of those that have been detected to contain
entities. Initially, the average precision of Stanford NER on this set reached 60\% for the German 
and 56\% for the English configuration of the entity extractor. This precision can be further increased 
by the simple post-processing, including two steps: 

\begin{itemize}
  \item[1.] Removal of the entities with long labels (i.e. the labels containing more than 2 terms). Our manual examination has shown that
  in many cases such long labels result from the extraction errors. 
  \item[2.] Removal of the entities that rarely occur in the URLs (the number of the URLs detected to contain the entity is less than 3 in the entire dataset).  
  According to our observations, entities extracted from very few URLs are often incorrect as opposed to the entities that 
  are repeatedly observed in URLs.  
\end{itemize}

After applying these corrections and re-evaluating the results, we observed the precision increase to 85\% for the German
and to 82\% for the English extractor on this dataset.

\subsection{Domain and Temporal Coverage of NER}
In this section we summarize the 
extraction results to better understand the domain coverage and the 
temporal coverage of the proposed method for this dataset.

Whereas some domains and domain categories possess entity-rich URLs, others do not. In addition, 
the number of URLs that contain entities in the Web archive can vary along the temporal dimension.
Therefore, in addition to the evaluation of the precision of the extraction method,
it is important to better understand the domain coverage of NER applied to the URLs in the Web archive. 
This analysis can help to better understand which parts of the Web archive 
can be made accessible using the proposed light-weight named entity annotation.

Overall 42,547,734 captures containing named entities have been identified by the extractor.
The frequencies of the named entities extracted from the URLs of these captures 
range from 2,301,917 to 3.
We decided a limit of 3 as a way to
maximize the NER precision, which is one of the post-processing steps described
above.
\begin{table}[!htb]
\vspace*{-5mm}
\label{tab:entities}
\begin{minipage}{.5\linewidth}

\begin{tabular}{|l|l|r|}
\hline
\textbf{Label} & \textbf{Type} & \textbf{Frequency} \\ \hline
deutschland           & location             & 2,301,917\\ \hline
berlin           & location                     & 628,300                   \\
\hline
hamburg           & location                     & 557,000                   \\
\hline nordrhein &       location               &      430,939             
\\\hline
muenchen            & location                     & 405,845                  
\\\hline
\end{tabular}
\end{minipage}%
\begin{minipage}{0.5\linewidth}
\hfill
\begin{tabular}{|l|l|r|}
\hline
\textbf{Label} & \textbf{Type} & \textbf{Frequency} \\
\hline michael jackson &       person               &      30,210 \\
\hline tommy hilfiger & person & 25,943 \\
\hline harald schmidt & person & 25,176              \\
\hline heidi klum & person & 21,291 \\ 
\hline merkel & person & 17,835 \\\hline
\end{tabular}
\end{minipage} 
\caption{The most frequent named entities of type ``location" and ``person" in
the urls of the
\GermanWeb dataset.}
\vspace*{-5mm}
\end{table}

The majority of the extracted entities are of type ``location'', followed by the type ``person''. 
The most frequent locations are local to Germany, whereas the person names are 
in many cases internationally known celebrities. Table~\ref{tab:entities} presents the most 
frequent entities of types ``location'' and ``person'' extracted from the 
\GermanWeb dataset and their frequencies (i.e. the number of captures). 

As an example of entities extracted from URLs, the Table \ref{tab:urls}
illustrates some URLs containing entities.
Up to two different entities could be extracted from those
URLs. Entities of type location as \textit{Berlin} and \textit{Prenzlauer-Berg}
were found in the same URL and persons as \textit{Franz Maget}
(german politician) and \textit{Katja Kessler} (german journalist).
\begin{table}[]
\centering
\caption{URLs containing entities}
\label{tab:urls}
\begin{tabular}{@{}ll@{}}
\toprule
{\bf URL} & {\bf Entities} \\ \midrule
\begin{tabular}[c]{@{}l@{}}http://www.hna.de/nachrichten/welt/costa-concordia\\
-zahl-vermissten-gestiegen-1565391.html\end{tabular} & Costa Concordia \\\hline
\begin{tabular}[c]{@{}l@{}}http://www.wg-gesucht.de:80/wohnungen-\\
in-Berlin-Prenzlauer-Berg.1529789.html\end{tabular} &
\begin{tabular}[c]{@{}l@{}}Berlin\\ Prenzlauer-Berg\end{tabular} \\\hline
\begin{tabular}[c]{@{}l@{}}http://www.stern.de:80/video/:Video-Franz-Maget\\
3A-Der-AuDFenseiter/638474.html?\end{tabular} & Franz Maget \\\hline
\begin{tabular}[c]{@{}l@{}}http://forum.gofeminin.de:80/forum/matern1
VERKAUFE-\\mein-DAS-MAMI-BUCH-katja-kessler.html\end{tabular} & Katja Kessler \\
\bottomrule
\end{tabular}
\end{table}
\subsubsection{Distribution of Entities by Domain Category}

\begin{figure}[htb!]
\vspace*{-10mm}
  \caption{Captures having entities within categories}
  \centering
    \includegraphics[width=1.0\textwidth]{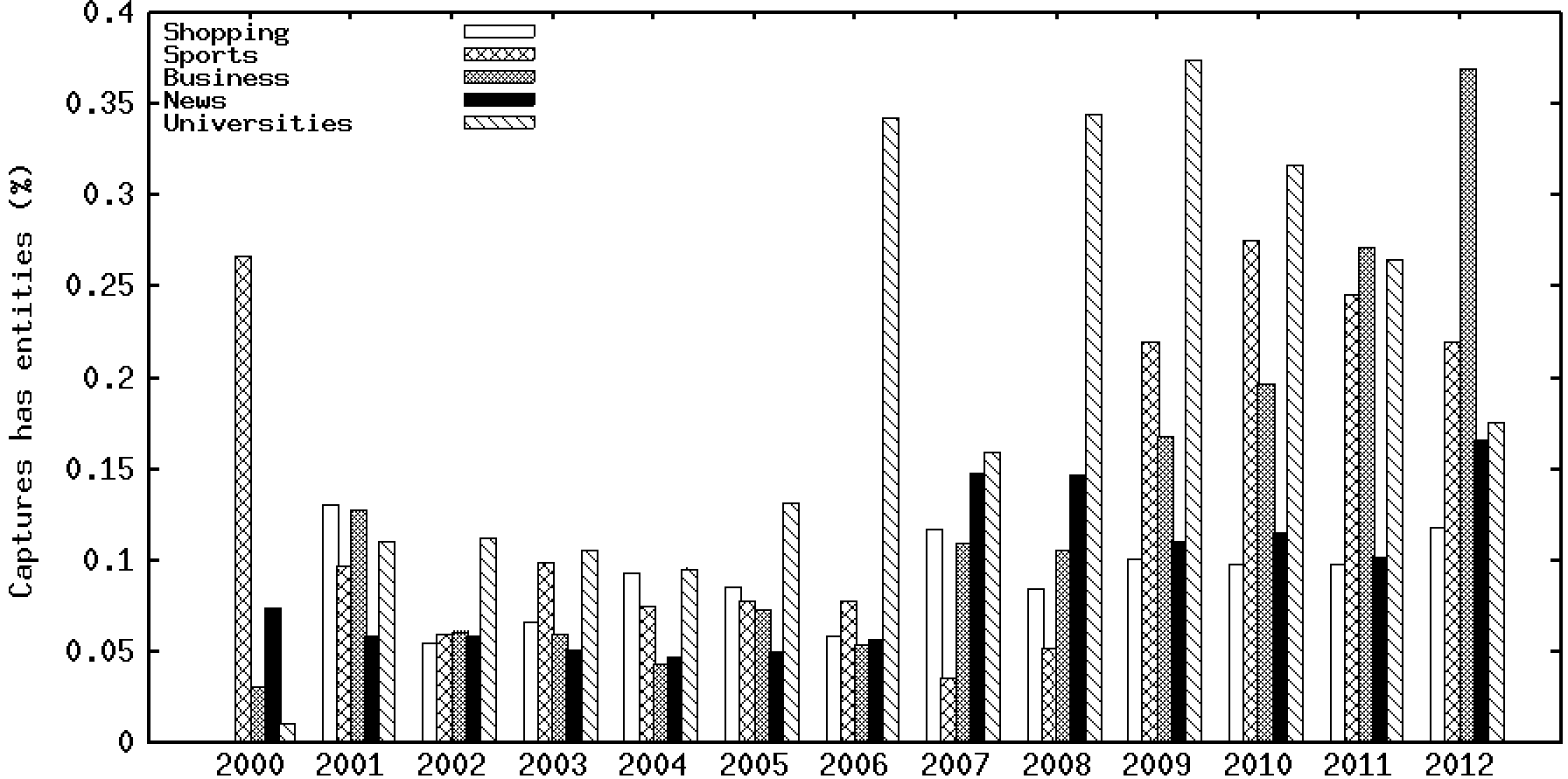}
    \label{fig:entities_percentage}
    \vspace*{-10mm}
\end{figure}

Fig.~\ref{fig:entities_percentage} illustrates the distribution of captures containing entities 
in several selected domain categories.

In this figure we normalized the
total number of entities by the total number of captures per year for each
specific category where X axis represent the years and Y axis the percentage of
captures with entities. 
We focus on those categories as they indicate interesting patterns with respect
 to the temporal dimension.
We can observe that between 2001 and 2006 the distribution 
remains overall stable while after 2006 several peaks can be observed. 

In the \textit{universities} category a first significant increase can be
observed between 2004 and 2005. By looking into the data it turns out that the domain \textit{uni-leipzig.de} dominates the crawl representing 19.81\% of the captures in 2005.
In 2006 the \textit{dblp.uni-trier.de}, a popular computer science bibliography,
dominates the captures regarding \textit{universities} representing 42.73\%. 

Continuing the analysis for \textit{universities}, in 2007 the same domain
 (\textit{dblp.uni-trier.de}) represents only 6.48\% of the total number of captures, and in that year all domains are more equally distributed. From
 2008 to 2011 the \textit{dict.tu-chemnitz.de} - a dictionary domain - dominates the collection and then gradually decreases
 from 33.26\% to 26.46\% of the total number of captures, respectively.
 In 2012 \textit{dblp.uni-trier.de} dominates again, but only represents 12.47\% of the captures. 
 
A first significant increase of entities in news can be observed in 2007. The
detailed analysis shows that in 2006 and 2007 the \textit{news} category was
dominated by \textit{spiegel.de} followed by \textit{openpr.de}, a press release portal.
 The portal uses the titles of the news articles as names for their html pages.
 Since the number of crawled 
 pages from this domain significantly increases from 200k pages in 2006 to 700k pages in 2007 we assume that the number of title-based URLs in the 
 archive also increased in this period. 

The \textit{sports} category shows a significant increase from 2007 to 2010 due
to a particular domain about transfer market of soccer
players - \textit {transfermarkt.de}.

Crawled pages from this domain increased from 500k in 2007 to 1.5
million in 2010, thus we expect that more players' names were mentioned and more
entities of type person and location were found. The quantity of captures
from such domain decreased in following years, therefore
we expect that less entities were extracted, as shown by the
graph.

Regarding \textit{business}, the majority of captures belong to
\textit{postbank.de} - a postal bank domain in Germany, which
increased from 680k in 2008 to 1.1 million in 2011 and it constitutes an
entity-rich domain for type location.

Overall Fig.~\ref{fig:entities_percentage} reveals a number of peaks from 2006 onwards. 
In the above example the reason was always the domination of a certain site. 
We assume that this is also the case for most of the other peaks that we observed
in previous analyses.

\subsubsection{Distribution of Entities by Type}

\begin{figure}[htb!]
    \vspace*{-8mm}
  \caption{Entities types retrieved through years}
  \centering
    \includegraphics[width=1\textwidth]{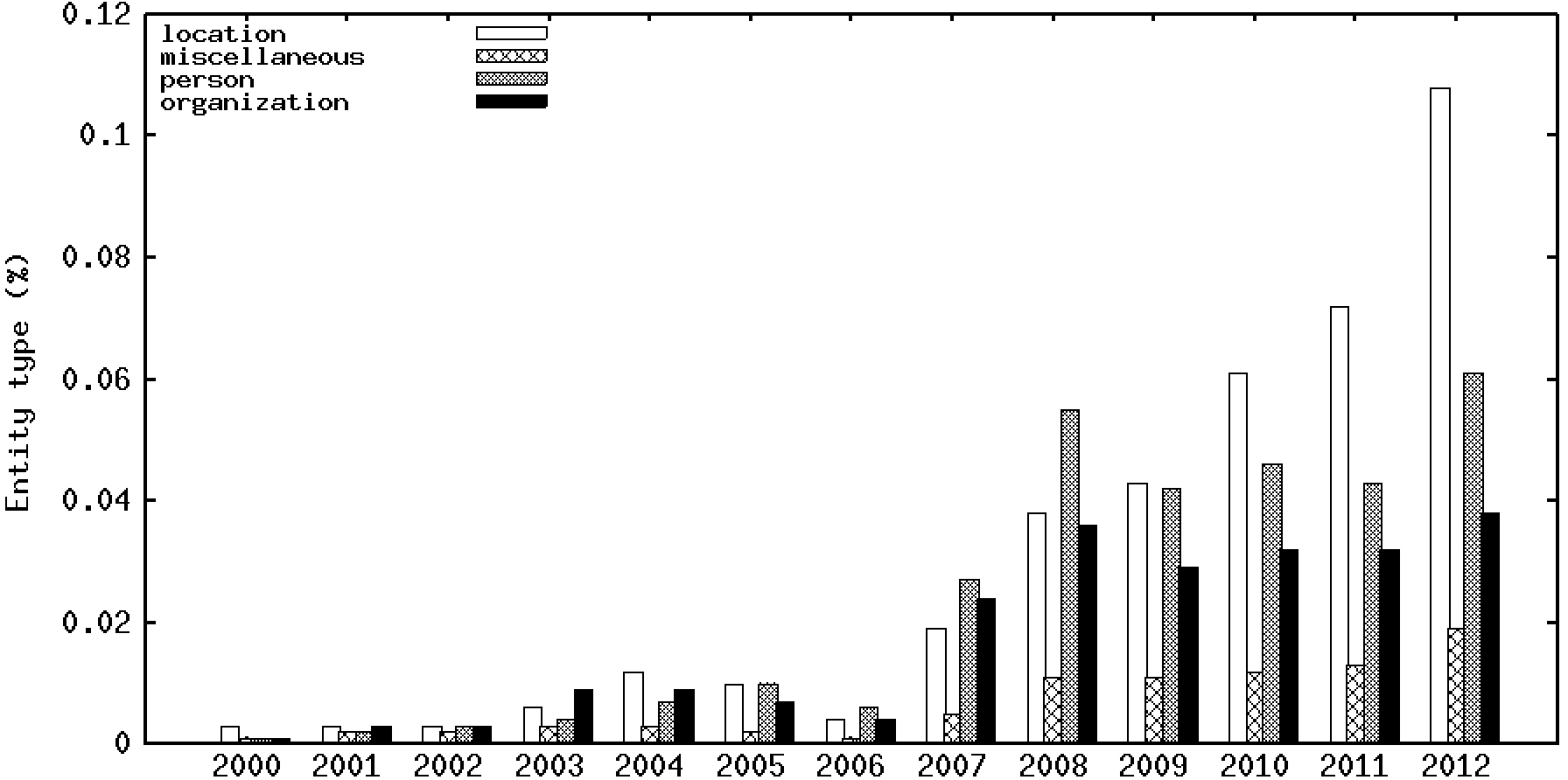}
    \label{fig:entity_type}
        \vspace*{-10mm}
\end{figure}

Fig. \ref{fig:entity_type} illustrates the distribution of entity types through
the years where the X axis represent the years and the Y axis the portion for each
entity type we extracted, normalized by the total number of entities.

As mentioned above, the most common entities are of type location followed
by the types person, and organization. In previous sections, we showed that while the
amount of captures significantly increased starting in 2006, many entity-rich
sites increased as well (e.g. \textit{postbank.de}, \textit{openpr.de},
\textit{transfermarkt.de}). Thus the number of entity types also increases,
as illustrated by Fig. \ref{fig:entity_type}, starting in 2006.
Since the majority of domains contains entity-rich URLs of type location, 
such type represent the most frequent one from all entities. 

%
%
%


\section{Related Work}
\label{sec:related}

The information contained in URLs has been analysed already in previous work,
however it is the first time that entities have been extracted using only the
information within URLs, the majority of current work rely on classification
tasks and do not consider entity extraction in URLs.
There are several works that classify the content of a document only based on its URL.
\citeauthor{Baykan:2011} conducted an extensive set of URL features and
classification methods to detect the topic of a Web document \cite{Baykan:2011}.
They report a precision of around $0.86$ and a recall between $0.36$ and $0.4$ on a multi-class topic classification using a combination of character n-grams of length $4$-$8$.
In their datasets they have about $32\%$ ``empty URLs'', i.e. URLs whose tokens or n-grams did not occur in the training set, which bounds the recall achievable using this data.
Similarly, \citeauthor{Kan:2005} discuss the applicability of URLs for general classification tasks \cite{Kan:2005}.
They also consider sequential features that take the typically hierarchical nature of URLs into account.

Special applications of URL classification are the detection of the document language \cite{Baykan:2013}, genre \cite{Abramson:2012} or locational relevance \cite{Anastacio:2009}.
In web crawling systems the information in URLs is used to detect duplicates \cite{Koppula:2010} or documents containing relevant types of information \cite{Hernandez:2012,Hernandez2014168}.
Furthermore, URLs have been used to detect malicious content \cite{Zhao:2013} as well as online advertising \cite{Raju:2012}.

A related field considers the anchor texts of links to a document.
Similarly to URL classification, this approach allows a performance comparable to content-based methods on many task at a lower processing cost.
For example, anchor-text based document ranking is significantly better for site finding tasks than a ranking using the content of the document \cite{Craswell:2001}.
However, these methods require the availability of anchor texts, whereas the URL of a document is always available.
Furthermore, anchor texts need to be extracted and collected across the document collection, which has a higher cost especially for large document collections.
Therefore we consider only URL based methods in this work.

\section{Discussion}
\label{sec:conclusion}

In this paper we presented our work on URL analytics towards providing 
efficient semantic annotations to large-scale Web archives.
Our results demonstrate that named entity recognition techniques can be effectively applied to URLs of the Web 
documents in order to provide an efficient way of initial document annotation.
Especially the years 2006 onwards provide useful information as the number of
URLs providing entities has increased ever since. For observing and analyzing longer periods, 
news and shopping domains turn out to be more useful, while dominating sites have a lower impact. 
In the future work we plan to further analyse term extraction techniques for the URLs
and to combine these techniques with light-weight content annotations to incrementally 
increase annotation coverage
while maintaining scalability and efficiency of the annotation process as well
as detecting temporal expressions still using the URL.

Understanding the dinamic of entities over time is important since we need to
know which part of our dataset has most promising entities. Therefore, when a
more specific search is needed, we should previously check which domains or
domains categories had the most entities and generate a subset on this specific domains, instead of considering the entire archive, which
sometimes is computationally unfeasible. The information we extracted from
URLs can be further analysed in the document content, as where in the HTML page (title, paragraphs, etc) the same entities or temporal expressions can be detected. This URL level analysis can support further keyword-based search algorithms in our web archive, also providing an overview of potential entity-rich domains (as those showed in last sections).

\section*{Acknowledgments}
\small{
This work was partially funded by the European Research Council under ALEXANDRIA (ERC 339233)
and the COST Action IC1302 (KEYSTONE). Tarcisio Souza is sponsored by a
scholarship from CNPq, a Brazilian government institution for scientific development. }

\bibliographystyle{plainnat}
\bibliography{keystone}
\end{document}